\documentclass[aps,pre,twocolumn,a4paper,footinbib,longbibliography,showpacs,groupedaddress,floatfix,superscriptaddress]{revtex4-2}% 

\usepackage[dvipsnames]{xcolor}
\definecolor{linkcolor}{rgb}{0.3,0.3,1.0} %hyperlink
\usepackage[pdftex,colorlinks=true, linkcolor= linkcolor, citecolor= linkcolor, urlcolor= linkcolor, hyperindex=true,hyperfigures=true]{hyperref} %hyperlink

\usepackage{amssymb,amsmath,amsfonts}
\usepackage{graphicx}
\usepackage{float}
\usepackage{bm}

\renewcommand{\vec}[1]{\bm{#1}}

\begin{document}

%%%
\title{Variational autoencoders understand knot topology
} 
%%%

%%%
\author{Anna Braghetto}
\affiliation{Department of Physics and Astronomy, University of Padova, 
Via Marzolo 8, I-35131 Padova, Italy}
\affiliation{INFN, Sezione di Padova, Via Marzolo 8, I-35131 Padova, Italy}

\author{Sumanta Kundu}
\affiliation{Scuola Internazionale Superiore di Studi Avanzati (SISSA), Via Bonomea 265, 34136 Trieste, Italy}
\affiliation{Dipartimento di Fisica, Università di Napoli Federico II, and INFN Napoli, Complesso Universitario di Monte
Sant’Angelo, 80126 Naples, Italy.}
%\affiliation{}

\author{Marco Baiesi}
\affiliation{Department of Physics and Astronomy, University of Padova, 
Via Marzolo 8, I-35131 Padova, Italy}
\affiliation{INFN, Sezione di Padova, Via Marzolo 8, I-35131 Padova, Italy}

\author{Enzo Orlandini}
\email{enzo.orlandini@unipd.it}
\affiliation{Department of Physics and Astronomy, University of Padova, 
Via Marzolo 8, I-35131 Padova, Italy}
\affiliation{INFN, Sezione di Padova, Via Marzolo 8, I-35131 Padova, Italy}

\begin{abstract}
Supervised machine learning (ML) methods are emerging as valid alternatives to standard mathematical methods for identifying knots in long, collapsed polymers. Here, we introduce a hybrid supervised/unsupervised ML approach for knot classification based on a variational autoencoder enhanced with a knot type classifier (VAEC). The neat organization of knots in its latent representation suggests that the VAEC, only based on an arbitrary labeling of three-dimensional configurations, has grasped complex topological concepts such as chirality, unknotting number, braid index, and the grouping in families such as achiral, torus, and twist knots. The understanding of topological concepts is confirmed by the ability of the VAEC to distinguish the chirality of knots $9_{42}$ and $10_{71}$ not used for its training and with a notoriously undetected chirality to standard tools. The well-organized latent space is also key for generating configurations with the decoder that reliably preserves the topology of the input ones. Our findings demonstrate the ability of a hybrid supervised-generative ML algorithm to capture different topological features of entangled filaments and to exploit this knowledge to faithfully reconstruct or produce new knotted configurations without simulations.
\end{abstract}

\maketitle

\section{Introduction}

Knots are topological states commonly observed in everyday life examples, such as disorderly stored garden hoses or headphone cables. In this case, knotted patterns are mostly detectable by simple visual inspection.  Knots can also occur at much smaller length scales where ropes and cables are replaced by biological macromolecules such as DNA~\cite{Rybenkov:1993:PNAS,Shaw:1993:Science:8475384,ArsuagaPNAS2002,ArsuagaPNAS2005,MarenduzzoPNAS2009, Marenduzzo:PNAS:2013} and proteins ~\cite{lua2006statistics,virnau2006intricate,jackson2017fold}. It is known that the abundance and complexity of knotted states depend strongly on the length and flexibility of the polymeric substrate~\cite{Sumners:Whittington:JPA:1988,Koniaris:prl:1991,orlandini:RMP:2007,Poier:Macromol:2014,Coronel:SoftMatt:2017:bending,orlandini2016local} as well as external conditions such as the quality of the solvent~\cite{Janse-van-Rensburg&Whittington:1990:J-Phys-A,Virnau:JACS:2005,Baiesi:2011:PRL}, the crowdness of the environment~\cite{dAdamo:Macromol:2015} and the geometry and degree of confinement~\cite{Micheletti:2006:J-Chem-Phys:16483240,Tubiana:PRL:2011,Micheletti:Macromol:2012,Micheletti:SoftMatt:2012,Liang_et_al_2012_Macro_Lett,Plesa_et_al_NatNato_2016,SumaE2991}. In turn, the presence of knots
influences the static and dynamic response of the host filament when subjected to stretching forces, extensional flows, or electric field~\cite{Farago:2002:EPL,Bao:PRL:2003,Micheletti_PhRep2011,zhang2019effects,Huang:JPCA:2007,Matthews:EPL:2010,Caraglio:PRL:2015,Narsimhan:MacroLett:2017:steady,Klotz:Macromol:2017:dynamics,Renner:SoftMatt:2015,Tang27092011,Rosa:PRL:2012}
Although identifying and classifying knotted states in polymeric chains is essential to fully comprehending their properties, this objective is far from trivial, especially when the chains are long and highly entangled in space.

\begin{figure}[t!]
 \centering
 \includegraphics[width=0.47\textwidth]{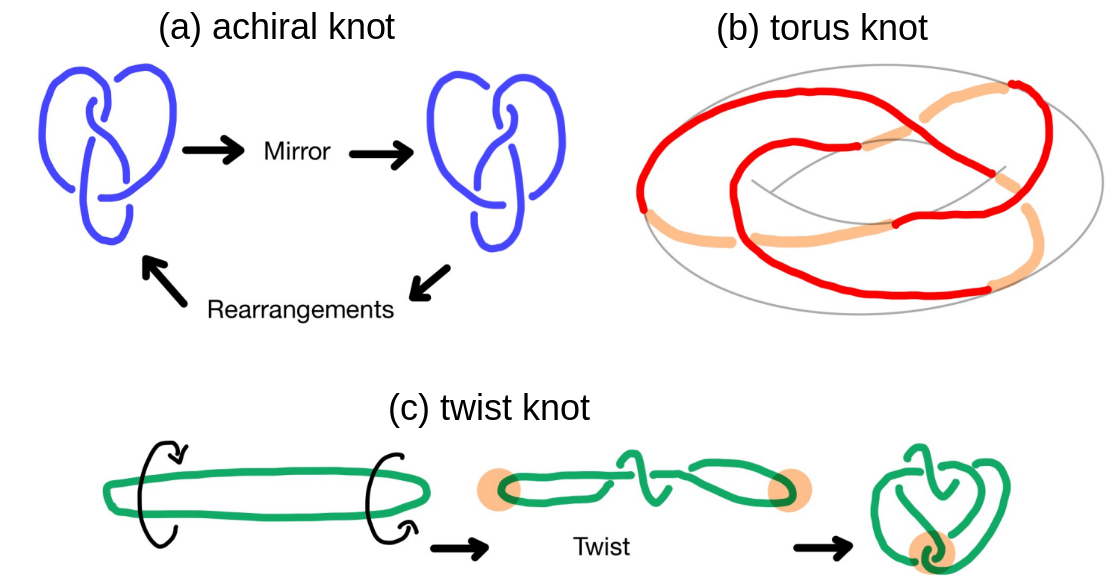}
 \caption{Sketches (a) of an achiral knot, (b) a torus knot, and (c) a twist knot. 
 } 
 \label{fig:families}
\end{figure}

Each knot is characterized by a list of properties that determine which families it belongs to. For example, there are achiral, torus, and twist knots among these families. Achiral knots do not have a positive or negative chirality, that is, their mirror images can be continuously mapped onto each other by preserving the knot type. This is the case for the knot $4_1$ sketched in Fig.~\ref{fig:families}(a).
Torus knots can be laid on the three-dimensional surface of a torus without loss of continuity, as in Fig.~\ref{fig:families}(b).
Finally, twist knots emerge from twisting an unknotted loop and then by linking its extremities, a process sketched in Fig.~\ref{fig:families}(c).

\begin{figure}[t!]
 \centering
 \includegraphics[width=0.3\textwidth]{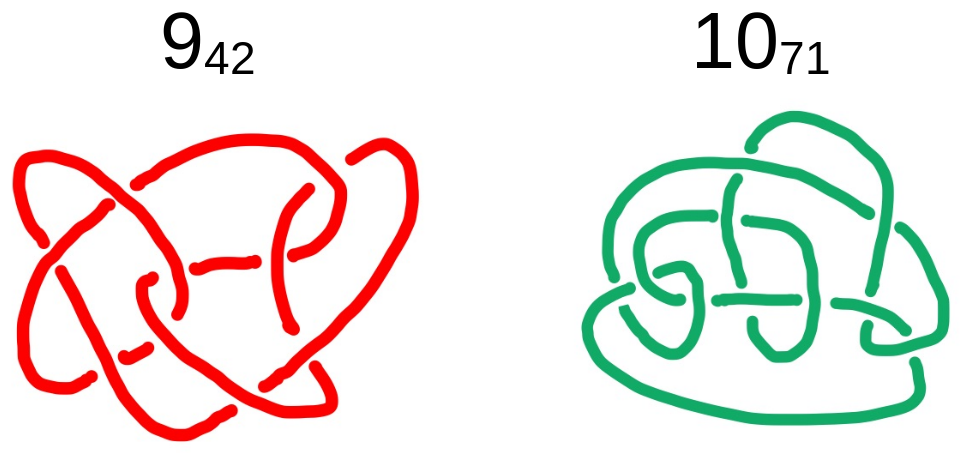}
 \caption{Two complex knots whose chirality cannot be computed with any known polynomial invariant.
 } 
 \label{fig:910}
\end{figure}

Traditionally, knot identification in polymeric systems has been based on mathematically rigorous projection methods followed by the computation of knot invariants such as Alexander, Jones, and HOMFLY polynomials~\cite{Adams:1994,orlandini:RMP:2007}. However, these approaches can be computationally intensive, especially for long polymers or complex three-dimensional configurations \cite{baiesi2012universal,Baiesi:2011:PRL,Micheletti:2006:J-Chem-Phys:16483240,Marenduzzo:PNAS:2013,Micheletti_PhRep2011}. Attempts to mitigate this issue through local deformations \cite{Koniaris&Muthukumar:1991a, Micheletti:2006:J-Chem-Phys:16483240, baiesi2014knotted} have shown promise, but remain time consuming and not universally applicable. Moreover, some knots such as the $9_{42}$ and the $10_{71}$ knots (see Fig.~\ref{fig:910}) are so complex that even powerful polynomial invariants (Jones and HOMFLY) cannot determine their chirality~\cite{Adams:1994,orlandini:RMP:2007}.

%%%%%%%%%%%%%%%%%%%%%%%%%%%
%%%%%%%%%%%%%%%%%%%%%%%%%%%

An attractive alternative is to abandon the mathematical rigor of the projection methods and identify knotted states as specific patterns hidden in severely entangled filaments. This is the ultimate goal of machine learning (ML) techniques, and in recent years, ML algorithms based on standard neural networks trained by supervised protocols
have emerged as an efficient methodology for knot identification~\cite{vandans2020,davies2021advancing,kauffman2020rectangular,jejjala2019deep,gukov2021learning,braghetto2023machine,sleiman2024geometric,wang2024integrating,bruno2024knots}.
Compared to projection methods, these ML algorithms, although not mathematically rigorous,
can distinguish knots with good accuracy also in situations where the polymeric substrate, being, for instance, severely confined, is characterized by a high degree of geometric entanglement~\cite{braghetto2023machine}.
Moreover, if the space of features accounts for geometrical properties of the polymer, such as the neighboring nonbonded monomers or the writhe, knot detection speed and accuracy can greatly increase with respect to standard projection methods~\cite{sleiman2024geometric,wang2024integrating}. 
A complete understanding of the mechanisms underlying the knot recognition process in an ML approach, including the structure of the internal representation learned by the algorithm, is not yet available.  Moreover, the possibility of generating new polymer configurations with a prescribed knotted state is still missing. 

Here, we introduce a hybrid approach to knot identification using variational autoencoders (VAEs), a generative ML technique that combines neural networks with probabilistic modeling. When a knot classifier based on its latent representation is introduced, the VAE becomes a discriminative tool.
We apply this VAE with classifier (VAEC) to a data set of confined, flexible ring polymers of varying lengths and topologies.

We find that the VAEC learns a well-structured latent representation of polymer configurations: Single knots are mapped to distinct regions, and the families described in Fig.~\ref{fig:families} emerge as sequences of knots of increasing complexity. Furthermore, the VAEC also sorts knots in the latent space depending on their unknotting number and, in another direction, on their braid index, as discussed below.
The fact that our VAEC has acquired some general way of classifying topology is confirmed from an out-of-sample application of a chiral-informed VAEC to complex knots never inserted in the training. These are the aforementioned $9_{42}$ and $10_{71}$ knots with chirality that is undetectable by rigorous mathematical methods. Despite the significant challenge, the VAEC distinguishes their opposite chiralities.

%%%%%%%%%%%%%%%%%%%%%%%%%%%
%%%%%%%%%%%%%%%%%%%%%%%%%%%

\section{Simulations and Neural Networks}\label{simulations_and_NN}

\subsection{Simulations}
\label{s:simul}

We use the same model as in Ref.~\cite{braghetto2023machine},
that is, we consider fully flexible knotted rings using a bead-spring model. Rings are collapsed due to confinement and include up to $N=512$ monomers. Hence, they present complex three-dimensional configurations that are highly challenging to ML techniques.

Each bead has a mass $m$ and a diameter $\sigma$.  To account for excluded volume interactions, each pair of beads interacts via the Weeks-Chandler-Anderson (WCA) potential with interaction strength $\epsilon$.
The subsequent beads along the chain are connected by the finitely extensible nonlinear elastic potential (FENE) with constants $k_0=30\sigma/\sigma^2$ and $R_0=1.6\sigma$. With this choice of the FENE and WCA parameters, we ensure that the topology of the initial configuration is preserved during its time evolution. The configurations of $N$ beads are spherically confined via a spherical indenter, that is, a force acting on each bead and pointing toward the center of the sphere. Using a sphere with radius $R$, we can simulate confined configurations with density $\rho=3N/(4\pi R^3)$. 
The set of Langevin equations that describe the dynamics of the system is numerically integrated in an NVE environment at temperature $T$ using the velocity Verlet algorithm implemented in the LAMMPS package~\cite{Plimpton:LAMMPS}. We set the integration time step $dt=0.001\tau$ where $\tau=\sigma\sqrt{m/\epsilon}$ is the characteristic simulation time (see more details in ~\cite{braghetto2023machine}). In the simulations, we choose to use $m=\sigma=T=\epsilon=1$, in dimensionless units.

%%%%%%%%%%%%%%%%%%%%%%%%%%%
\begin{figure}[t!]
 \centering
 \includegraphics[width=0.47\textwidth]{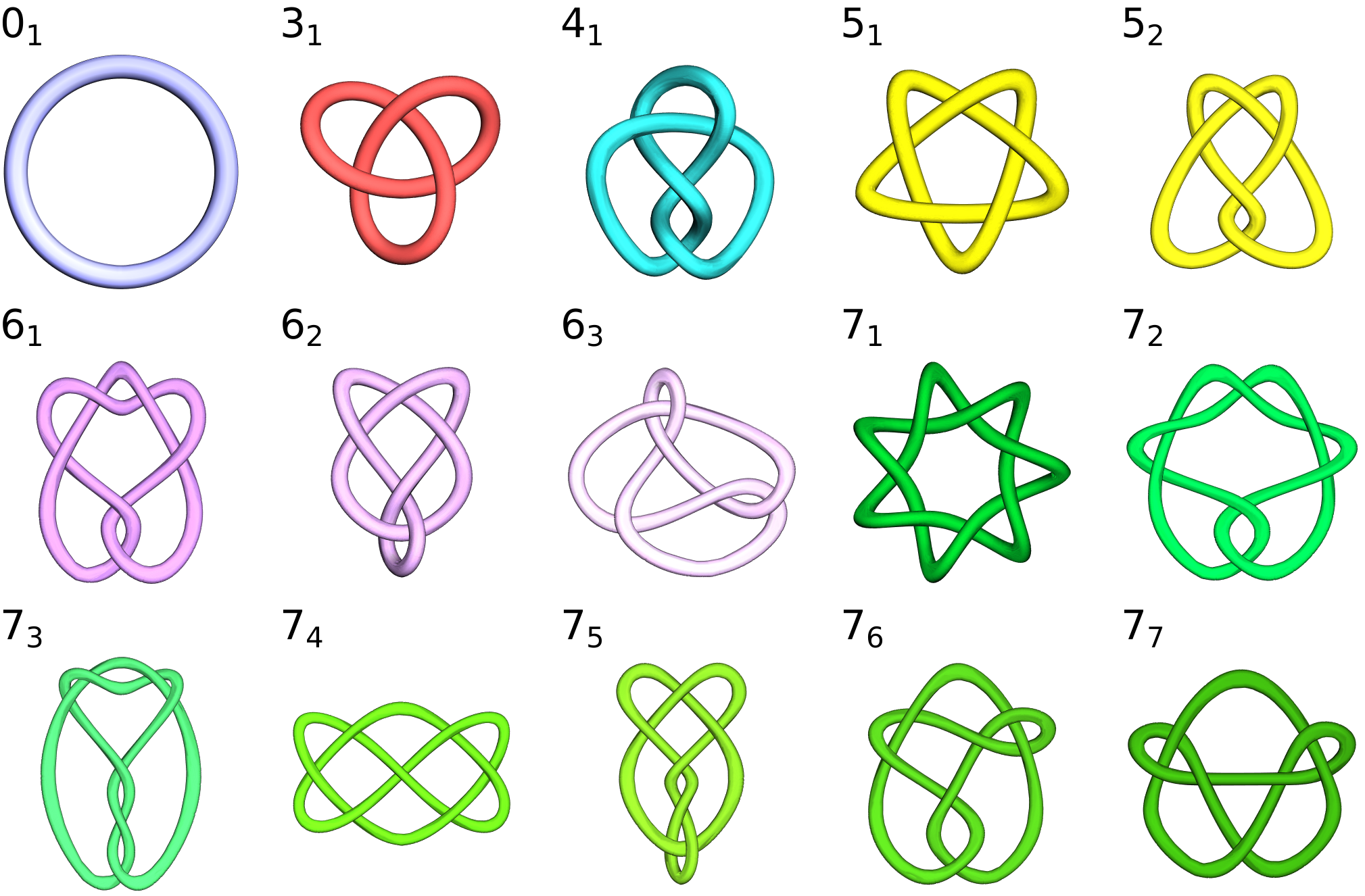}
 \caption{Table of knots used for the training of the VAEC. 
 } 
 \label{fig:table}
\end{figure} 
%%%%%%%%%%%%%%%%%%%%%%%%%%%

In this paper, we consider ring chains with fifteen different topologies;  according to the minimal crossing representation, these are the unknot $0_1$, and the trefoil knot $3_1$, the figure eight knot, $4_1$, the five crossing knots $5_1$ and $5_2$, the six crossing knots $6_1$, $6_2$, $6_3$, and the seven crossing knots $7_1,\ldots,7_7$. Their basic topology is recalled in Fig.~\ref{fig:table}. For each knot type, we generate confined configurations for three different chain lengths $N=128, 256$, and $512$. 
For each $N$, the value of $R$ is chosen to have confined polymers with a monomer density $\rho_1=0.07$ or higher. With $\sigma=1$, the corresponding volume density of the monomers is $\approx \rho_1/2$.

\subsection{Variational Autoencoders with Classifier}

The variational autoencoder (VAE) is a probabilistic generative neural network~\cite{Kingma_Welling_2014,rezende2014stochastic,cinelli2021variational}. It can be used for dimensionality reduction and feature extraction. This model compresses data into a lower-dimensional space (thanks to its bottleneck architecture) and also captures the underlying probabilistic distributions of the data. This characteristic enables a VAE to generate synthetic data that closely resembles the original data set. As a result, VAEs have found applications in various fields, including image processing, natural language processing, and generative modeling, making them a powerful tool in the realm of machine learning. 

%%%%%%%%%%%%%%%%%%%%%%%%%%%
\begin{figure}[t!]
 \centering

 \includegraphics[width=0.47\textwidth]{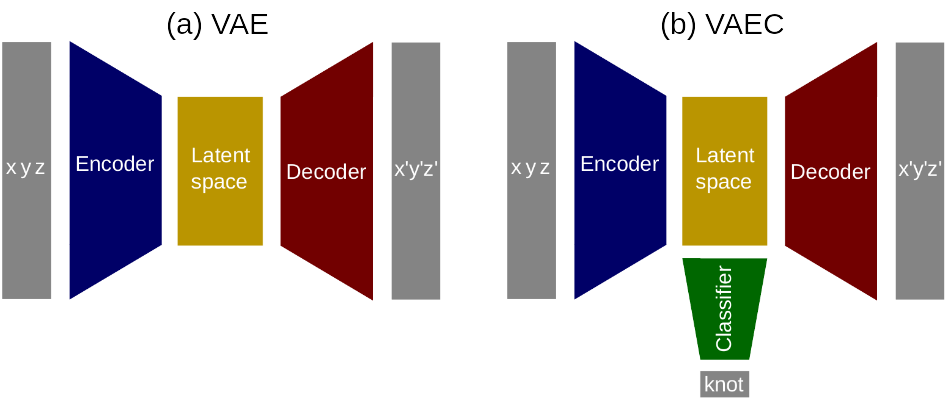}
 \caption{Sketches (a) of the standard VAE architecture and (b) of the VAEC implemented in this study. The input to the encoder consists of a set of coordinates $\vec x = (x_i,y_i,z_i)$ for $ i\le N$, representing a polymer configuration. These coordinates are mapped into a latent space by the encoder. Such a low-dimensional $\vec{z}$ representation is re-expanded to three-dimensional ring chains by the decoder. In addition, in the VAEC, the latent representation of the polymer serves as an input for the classifier designed to predict the topology of the polymer. 
 } 
 \label{fig:VAEC}
\end{figure} 
%%%%%%%%%%%%%%%%%%%%%%%%%%%

The power of VAEs comes from their capability to process the information summarized in the low-dimensional latent space. As sketched in Fig.~\ref{fig:VAEC}(a), they contain an encoder function ($F$) that takes an input sample ($\vec{x}$, in our case the list of coordinates $(x_i,y_i,z_i)$ with $1\le i\le N$) and maps it to a latent representation $\vec z = F(\vec{x})$ with $32$ components.
Then, the original $3N$-dimensional data are reconstructed by a decoder function $D$. Ideally, this generated sample $\vec{x'} = D(F(\vec{x})) = D(\vec{z})$ should closely resemble the original input $\vec x$. The loss function $\mathcal{L}_1$ associated with the VAEs comprises two distinct terms, 
\begin{equation}\label{eq:vae_loss}
    \mathcal{L}_1(\vec{x},F,D) = |\vec{x}-D(F(\vec{x}))|^2 + \mathrm{KL}(\mathcal{N}(\vec 0, \vec 1), F(\vec{x}))
\end{equation}
The first term represents the squared distance between the original input $\vec{x}$ and the output generated by the decoder, $D(\vec{z})$. This term plays a crucial role during the training process as it aims to minimize the discrepancy between the input and the decoder's output. The second term is the Kullback-Leibler divergence, which quantifies the difference between the latent (probabilistic) representation $F(\vec{x})$ and a specified reference distribution, in this case a standard normal distribution
$\mathcal{N}(\vec 0, \vec 1)$ in the $\vec z$ space. This Kullback-Leibler divergence term is essential for training a VAE, as it encourages the model to learn to sample appropriately from the latent space, thereby enhancing its utility as a generative tool.
In summary, the VAE framework combines these two components in its loss function to effectively balance reconstruction accuracy with the ability to generate new samples from the learned latent space.

When data are organized into distinct clusters, related clusters should emerge in the latent space during the training process. Thus, a low-dimensional visualization of $\vec{z}$ might highlight specific features of the data and reveal the underlying structures. However, knotted polymer configurations exhibit a high degree of variability. It might be difficult to distinguish clusters in $\vec{z}$ because geometrically similar configurations, despite their knot type, are mapped to nearby $\vec{z}$ points. Due to this mixing in $\vec{z}$ of configurations with different topologies, our attempts to apply standard VAEs have not produced any discernible clusters related to knots within the latent space. Therefore, the latent representation of the VAE does not help to recognize the knot type.

The latent representation becomes organized according to topology once we transform a VAE into a supervised machine learning model by incorporating a classifier ($C$) that analyzes the latent space, as illustrated in Fig.~\ref{fig:VAEC}(b). Since our main objective is to generate configurations with a known topology, we assigned the classifier the task of predicting the knot type of the input configuration, that is, the output generated by the classifier $C(\vec{z})=C(F(\mathbf{x}))$ should determine the knot category. The three components of the model, encoder, decoder, and classifier, are trained concurrently.
To facilitate this process, we modify the VAE loss function as 
\begin{equation}\label{eq:VAEC_loss} \mathcal{L}_2(\vec{x},F,D) = \mathcal{L}_1(\vec{x},F,D) + \sum_k y_k \log C(F(\vec{x}))_k \end{equation}
to create what we term a VAE classifier (VAEC). This modified loss function incorporates an additional term that represents the categorical cross entropy between the known knot type for a sample with knot type $k$ (for which $y_k=1$ and $y_{k'}=0$ for $k'\ne k$) and the output generated by the classifier. 

This training approach imposes a significant constraint on the structure of the latent space, particularly when using a relatively simple classifier. As a result, distinct clusters begin to form in the $\vec{z}$ space, driven by the need to classify knot types solely based on the information encoded there. These clusters predominantly consist of knots of the same type and, as elaborated in the following discussion, similar features of knots tend to characterize neighboring clusters.

\subsection{Architecture}

We carefully select the model architectures for our VAEC by keeping our primary objectives in mind: the prediction and generation of polymer configurations, particularly about knot types. To effectively handle sequential data, we opt for a transformer architecture for the encoder~\cite{vaswani2017attention,lin2022survey}. This choice enables us to capture the intricate dependencies within the data.
For the task of predicting the knot type, we implement a small fully connected feedforward neural network as our classifier. This design allows for robust learning of the relationships between the encoded representations and the corresponding knot types.
In the decoding phase, the main purpose is to increase the sample from the latent dimension to the polymer length, that is, to reconstruct the polymer structures from the latent representation $\vec z$. To this end, we use transpose convolutional layers, which generate simplified representations of polymer configurations. 

\begin{figure}[t!]
 \centering
\includegraphics[width=.47\textwidth]{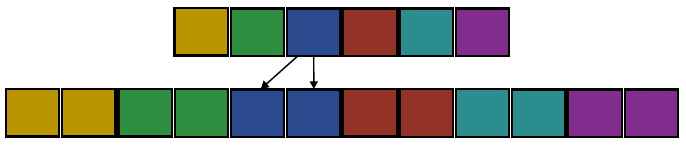}
 \caption{Kernel manipulation to make the model suitable for chains longer than $N$. Each filter is doubled for chains of length $2N$. Iterating this step, we deal with chains of length $4N$.}  
 \label{fig:kernel}
\end{figure}

To enhance the flexibility and adaptability of the model, we also incorporate a preliminary convolutional layer into the encoder. This layer processes the coordinate sequences using a set of learnable filters.
They are represented as colored squares at the top of Fig.~\ref{fig:kernel}.
They allow for easy tuning of the input if the chain length is doubled. In that case, we simply double the size of the trained filters, as sketched in Fig.~\ref{fig:kernel}.
This can be iterated, which allows us to analyze chains with $N=256$ and $N=512$ even if the model was trained with $N=128$. Hence, our architecture is well-suited to handle a diverse range of polymer lengths.

%%%%%%%%%%%%%%%%%%%%%%
%%%%%%%%%%%%%%%%%%%%%%

\section{Validation}

%\subsection{Validation on training length $N=128$}

\subsection{Simple knots}

\begin{figure}[t!]
 \centering

\includegraphics[width=.3\textwidth]{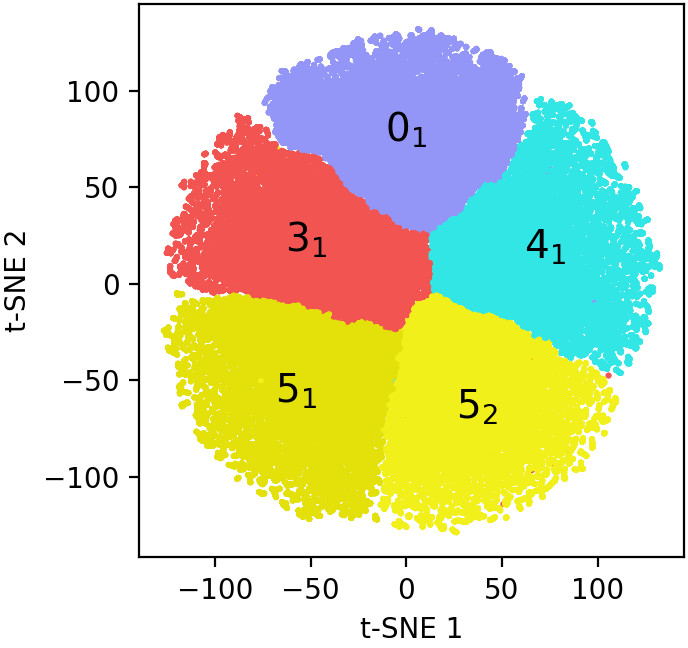}
 \caption{Two-dimensional projection with t-SNE of the latent representation for the VAEC trained on rings at density $\rho = \rho_1 = 0.07$ including knots up to five crossings. Data points for different knots have different colors and cluster in specific areas. Since the mapping of t-SNE is rotationally invariant, we will always position the $0_1$ on top of the diagram for better readability.}  
 \label{fig:tSNE1}
\end{figure}

To start, we test the VAEC performances with knots up to five crossings ($0_1,\ldots,5_2$). 
We train the encoder with polymer configurations with $N=128$ beads and compressed at a monomer number density $\rho_1=0.07$.
First, we validate its accuracy for chains of the same length. The resulting encoding in the latent space is illustrated in Fig.~\ref{fig:tSNE1}, where we map the latent representations to a two-dimensional one with the t-SNE algorithm~\cite{mehta2019high}.
The latent representation clusters configurations with the same knot type. Thus, the VAEC successfully learns to identify the knot type, as we have integrated a classifier that operates on the latent vectors. 
One can also note some ordering in the $\vec{z}$ space, with the simpler unknot $0_1$ on the opposite side of the more complex $5_1$ and $5_2$ knots. More order will emerge by analyzing more complex knots.

%%%%%%%%%%%%%%%%%%%%%
%%%%%%%%%%%%%%%%%%%%%

%%%%%%%%%%%%%%%%%%%%
 \begin{figure*}[t!]
 \centering
 \includegraphics[width=.75\textwidth]{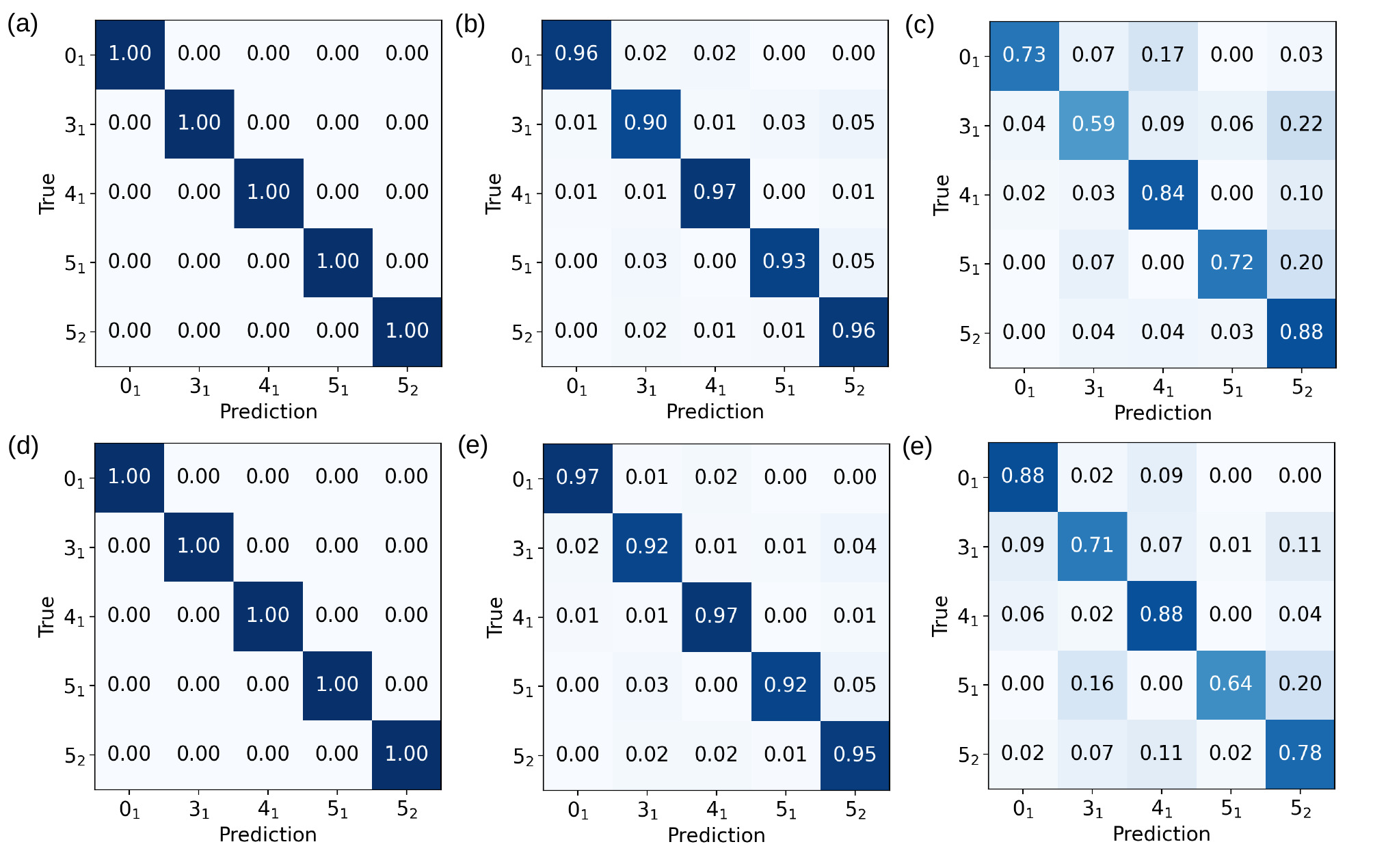}
 \caption{(a)-(c) Confusion matrix for several validation sets, using the always a training set of configurations with length $N=128$ and density $\rho_1$: (a) for polymers with the same length ($N=128$) of those used for training, (b) for $N=256$ and (c) $N=512$. Panels (d)-(f) refer to the case in which training was still carried out on polymers of length $N=128$  but now sampled at three different densities $\rho_1,\rho_2,\rho_3$: (d) $N=128$, (e) for $N=256$ and (f) $N=512$.}
 \label{fig:conf1}
%\end{figure} 
%%%%%%%%%%%%%%%%%%%%
%%%%%%%%%%%%%%%%%%%%%
%\begin{figure*}[t!]
 \centering
 \includegraphics[width=.8\textwidth]{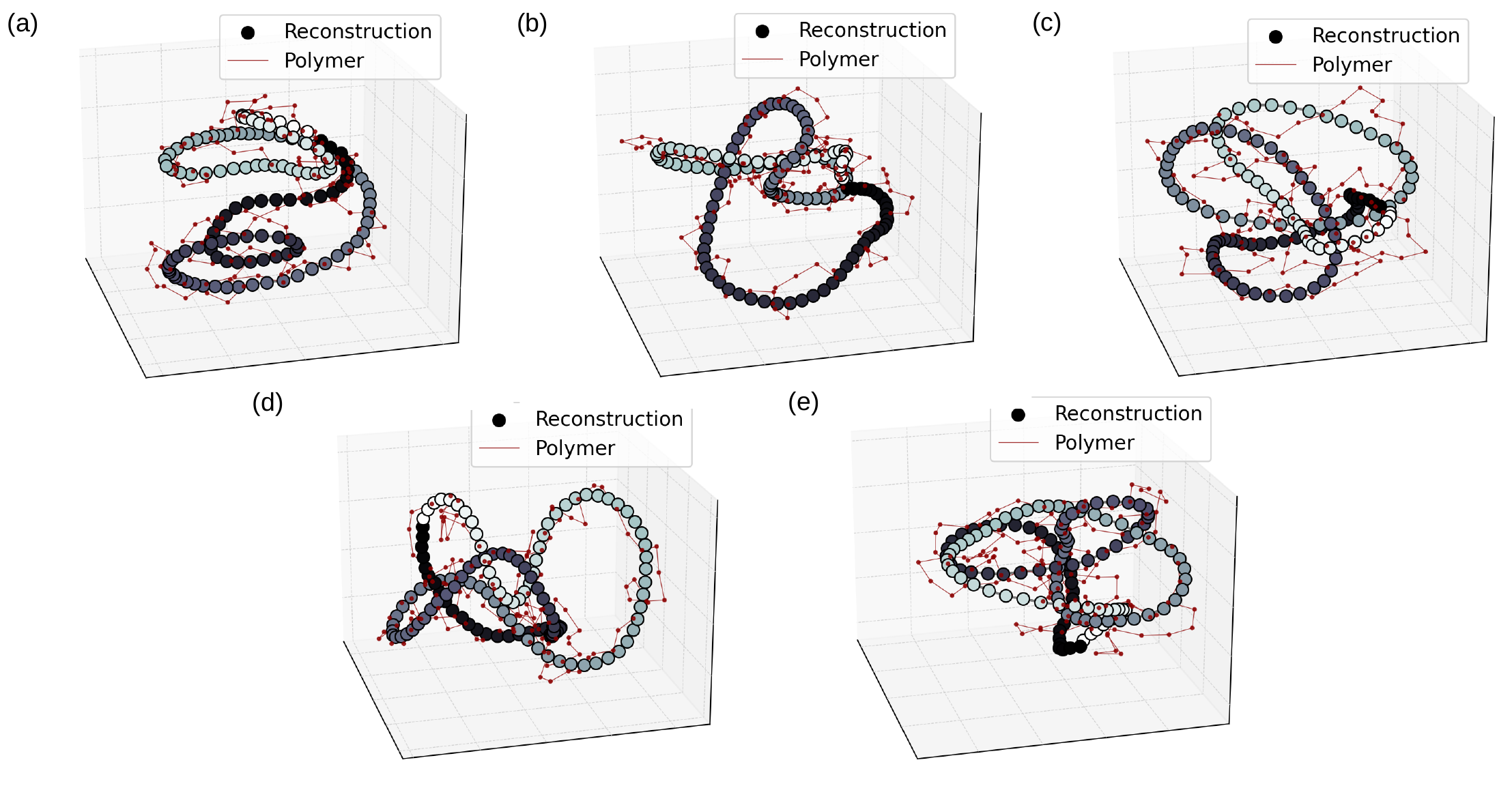}
 \caption{Examples of original configurations (thin lines and small points) and reconstructed polymer configurations (bigger dots) for different knots: (a) $0_1$, (b) $3_1$, (c) $4_1$, (d) $5_1$, and (e) $5_2$. 
}  
 \label{fig:fig3}
%\end{figure*} 
% \begin{figure*}[t!]
 \centering
 \includegraphics[width=.99\textwidth]{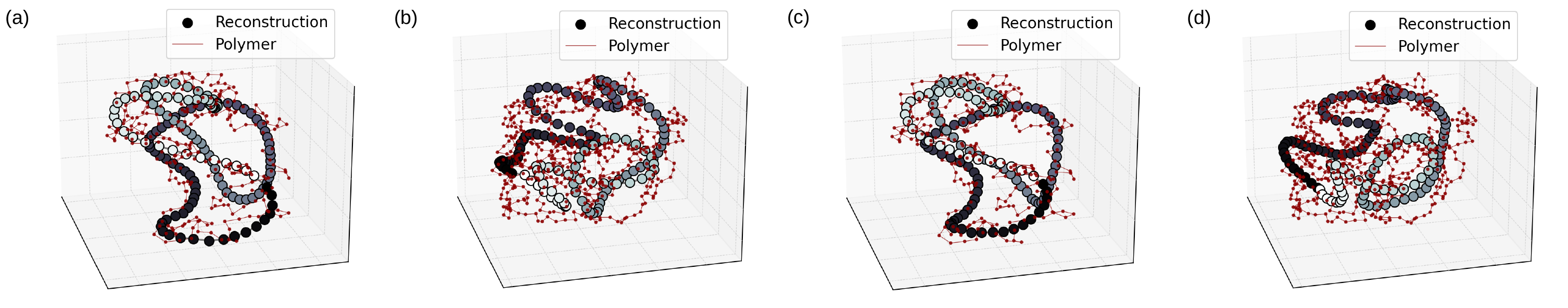}
 \caption{Reconstruction for model trained on the single number density $\rho_1=0.07$ for (a) $N=256$ and (b) $N=512$, and for model trained on multiple densities $\rho_1,\rho_2,\rho_3$ for (c) $N=256$ and (d) $N=512$. 
 } 
 \label{fig:fig7}
\end{figure*} 
%%%%%%%%%%%%%%%%%%%%%%%%%%%%%%%%%%%%%%%%%

Using latent representations, we predict the type of knot using a classifier, achieving an excellent accuracy of 99.8\% (see also Fig.~\ref{fig:conf1}(a)). In addition to classifying the knots, we leverage these latent representations to reconstruct the polymer configurations. The application of a convolutional filter in the encoder and of transpose convolutional filters~\cite{dumoulin2016guide} in the decoder results in reconstructed configurations, as shown in Fig.~\ref{fig:fig3}, which exhibit a smoother and simpler appearance compared to the original configurations.

\subsection{Generalization on longer chains}

Knots in globular polymers are delocalized~\cite{Baiesi:2011:PRL,baiesi2014knotted}. Hence, we assume that the patterns with $m$ monomers characterizing the knot in chains with $N=128$ are most likely mapped to the patterns with $2m$ monomers for chains with $N=256$. Applying double-length convolutional filters to these patterns for $N=256$ should produce a similar signal in the neural network. This logic leads to a double-length first convolutional layer of the encoder as illustrated in Fig.~\ref{fig:kernel}: each weight is copied twice in the double-length filter. This weight manipulation adapts the model to predict the properties of longer polymer chains. This is achieved while maintaining a fixed output shape of $N=128$ beads for reconstruction. As a side product of this method, this fixed output length produces a simplified version of long chains, as illustrated in Fig.~\ref{fig:fig7}.

However, let us focus on the accuracy of the VAEC prediction of knot classes in polymer chains longer than $N=128$.
The validation results for the training size $N=128$ at a density of $\rho=\rho_1=0.07$, are reported in terms of a confusion matrix in Fig.~\ref{fig:conf1}(b) for $N=256$ and Fig.~\ref{fig:conf1}(c) for $N=512$.
As expected, the performance tends to degrade for long polymers. However, the accuracies remain quite good even for $N=512$.

We find similar results if the training at $N=128$ is performed on a broader data set that includes various densities, specifically
$\rho=[\rho_1,\rho_2,\rho_3]=[0.07,0.14,0.28]$, see Fig.~\ref{fig:conf1}(d)-(f)). 
These results confirm the VAEC's capability to generalize and obtain good accuracy also for a test performed on chains four times longer than those used for training.

%%%%%%%%%%%%%%%%%%%%%%%%%%%%%%%%%%%%%%%%%
\begin{figure}[t!]
 \centering
 \includegraphics[width=.48\textwidth]{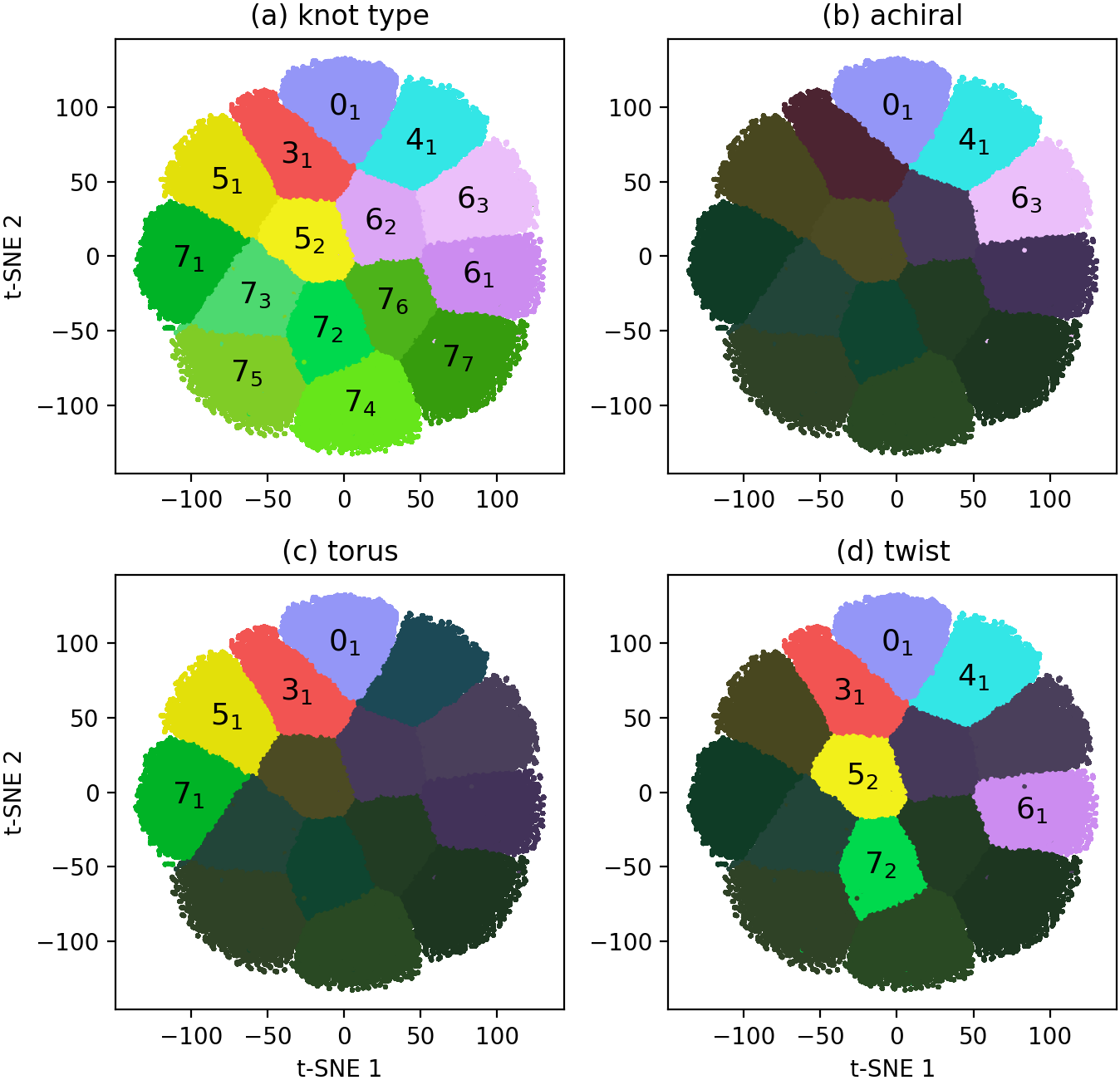}
 \caption{(a) Embedding to two dimensions of the latent representation for knots up to seven crossings, displaying clusters of knots with the same crossing number (different shades of the same color) and a gradient of complexity from the simplest knot ($0_1$ on top) to more complex ones. (b) The family of achiral knots emerges as a sequence of increasing complexity. (c) Similarly, the family of torus knots forms a line starting from the simpler one and ending with the more complex $7_1$ knot. (d) Also twist knots mostly form a cluster in the latent space.
 } 
 \label{fig:tSNE2}
\end{figure} 
%%%%%%%%%%%%%%%%%%%%%%%%%%%%%%%%%%%%%%%%%
\subsection{Knots up to seven crossings}

Next, we test the VAEC's ability to classify more complex knots and sort their families in the latent space. The latent space, as a result of the classifier, reveals a structured representation of information regarding different knot types.

%%%%%%%%%%%%%%%%%%%%%%%%%%%%%%%%%%%%%%%%%
 \begin{figure}[t!]
 \centering
 \includegraphics[width=.47\textwidth]{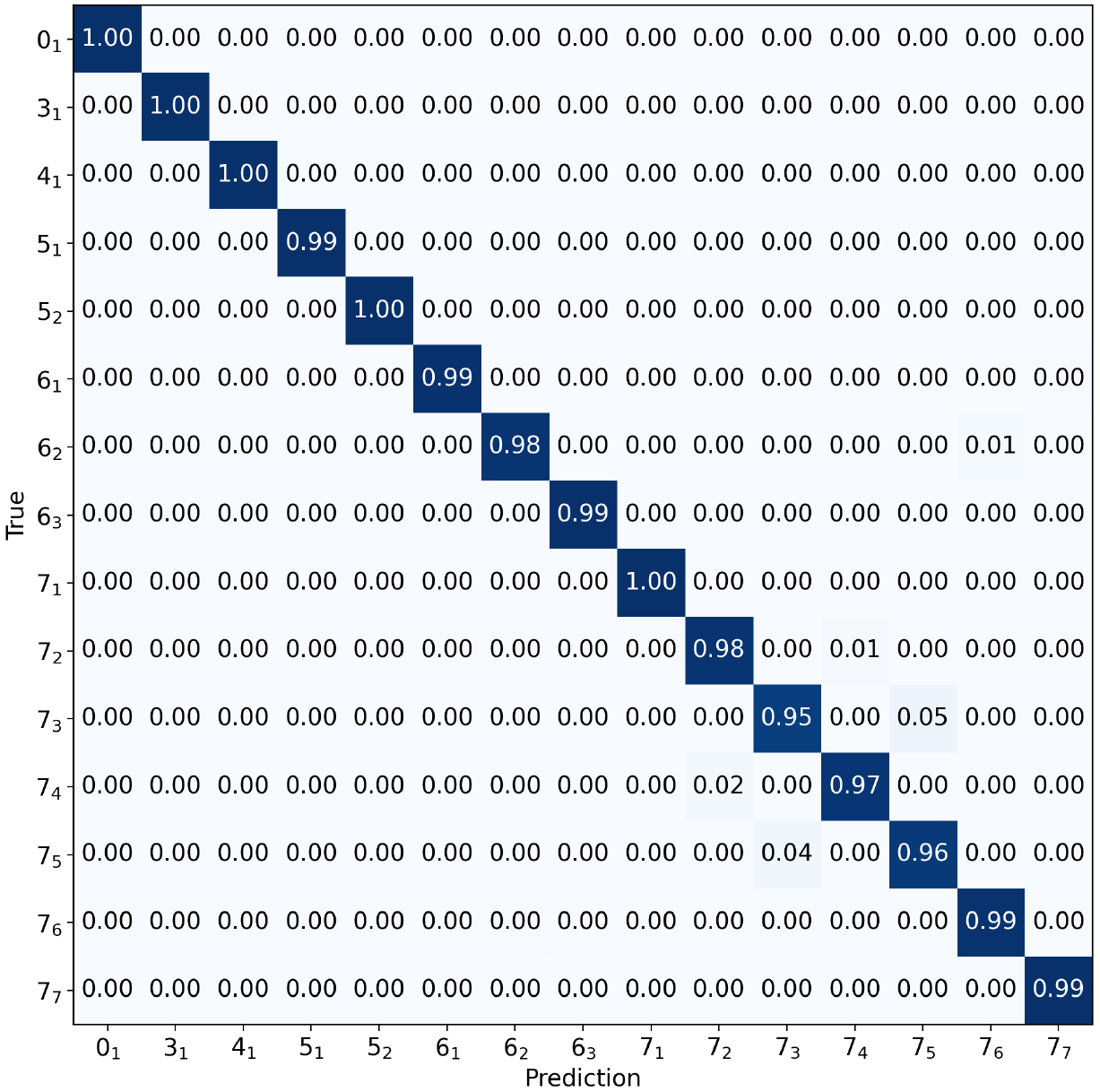}
 \caption{(a) Confusion matrix for VAEC trained with knots up to seven crossings.
 }
 \label{fig:conf15}
\end{figure} 
%%%%%%%%%%%%%%%%%%%%%%%%%%%%%%%%%%%%%%%%%
As before, we train the VAEC with a data set of $N=128$ long configurations compressed at number density $\rho_1$=0.07, but now we include knots with up to seven crossings (see Fig.~\ref{fig:table}). In Figure~\ref{fig:tSNE2}(a), we observe a neatly sorted latent representation of the knotted rings: It consists of 15 distinct clusters, each corresponding to a specific knot type. Furthermore, we note the presence of a sequence of achiral knots of increasing complexity (Fig.~\ref{fig:tSNE2}(b)), torus knots of increasing complexity (Fig.~\ref{fig:tSNE2}(c)), and a cluster of twist knots (Fig.~\ref{fig:tSNE2}(d)). 
Given the latent representations, we evaluate the excellent performance of the classifier utilizing the confusion matrix in Fig.~\ref{fig:conf15}.

\section{Including chirality}

\subsection{Knot families}

To achieve a better understanding of how knots are embedded within the latent space and to generate more specific knot types, we decided to train a model that incorporates the notion of \emph{knot chirality}. This is done by increasing the number of output units of the classifier to effectively distinguish between knots of different chirality.
More precisely, we separate the unit designated for the knot type $3_1$ into two distinct units: one for the left-handed chiral knot $3_{1}^-$ and another for the right-handed chiral knot $3_{1}^+$. Since $12$ out of the $15$ considered knot types are chiral, the total number of classes increases to $27$.

%%%%%%%%%%%%%%%%%%%%%%%%%%%%%%%%%%%%%%%%%
 \begin{figure}[t!]
 \centering
 \includegraphics[width=.48\textwidth]{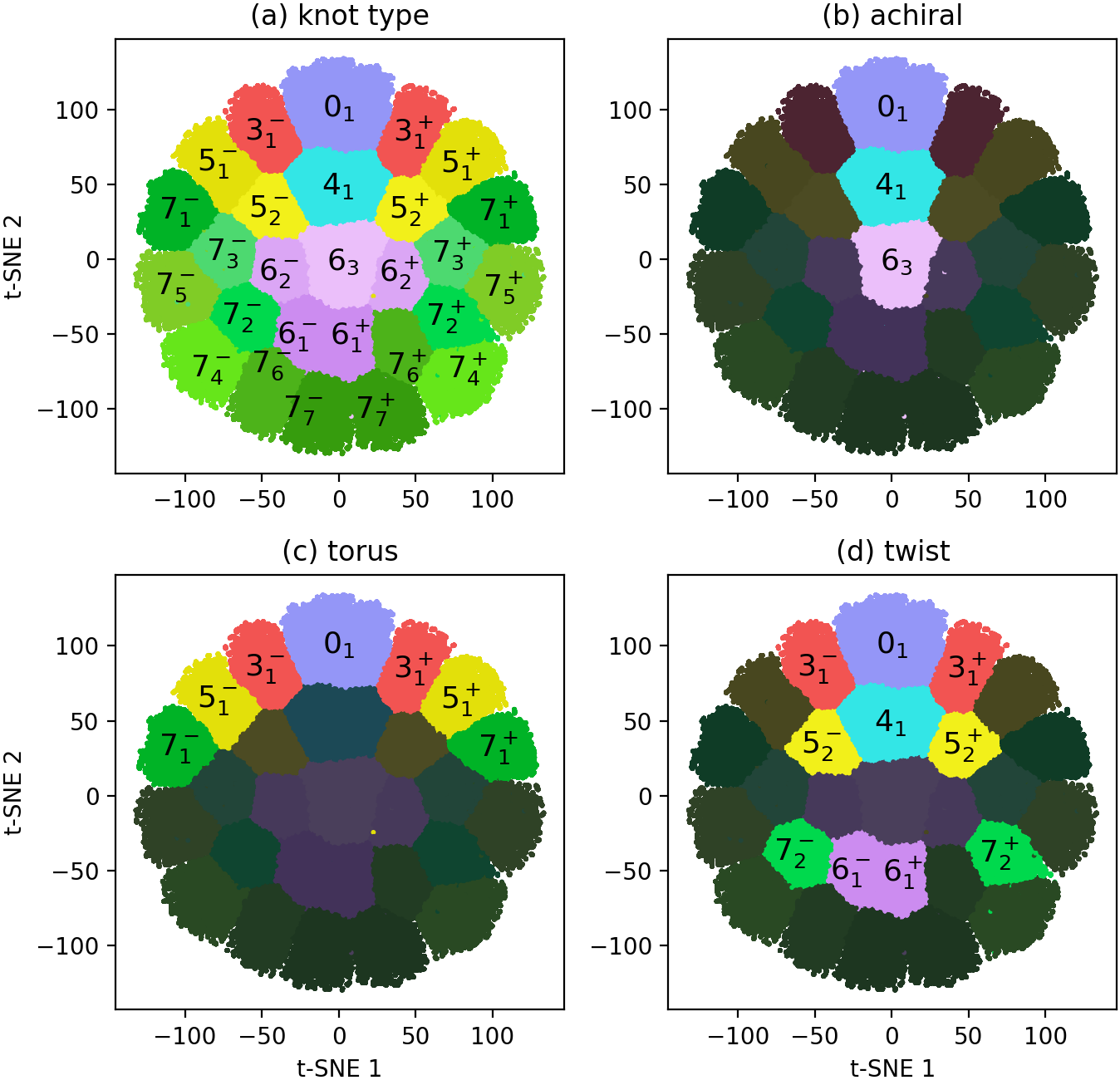}
 \caption{As in Fig.~\ref{fig:tSNE2}, but including knot classification also according to chirality.
} 
 \label{fig:tSNE3}
\end{figure} 
%%%%%%%%%%%%%%%%%%%%%%%%%%%%%%%%%%%%%%%%%

To create a data set informed by the chiral properties of the available configurations, we apply a coordinate transformation to half of the configurations that exhibit chiral topology. Specifically, we consider the mapping $\vec{x}=(x_i,y_i,z_i)\rightarrow \vec{x}^*=(- y_i, z_i,x_i)$.
We choose this transformation to better decorrelate the samples within our dataset. A simpler transformation, such as flipping a single coordinate $(x_i,y_i,z_i)\rightarrow (- x_i, y_i,z_i)$, would change the configuration's chirality, but it might also allow the model to learn only the change of sign of that coordinate. Finally, to decrease the possibility that the model would
learn correlations between configurations, we apply the transformation only to configurations belonging to the second half of the simulated time series.

The results of this procedure are applied to the validation set, which includes both chiral and achiral knots of the training set, as illustrated in Fig.~\ref{fig:tSNE3}(a). Concerning chirality, we observe a symmetric representation of the knots in the latent space (somewhat a symmetrized version of that in Fig.~\ref{fig:tSNE2}(a)): achiral knots are positioned at the center of the latent structure (see also Fig.~\ref{fig:tSNE3}(b)), while negative and positive chiral knots are located on the left and right of the latent space, respectively. In addition to this emerging chirality structure, we find that the model neatly organizes knot families, positioning torus knots at a margin of the latent space (Fig.~\ref{fig:tSNE3}(c)) and twist knots toward its top center (Fig.~\ref{fig:tSNE3}(d)). However, the clustering of twist knots is not perfect. In the following, we highlight other trends that the VAEC has generated in the latent space, possibly incompatible with a nice clustering of twist knots.

\subsection{Knot complexity}

\begin{figure}[t!]
 \centering
 \includegraphics[width=.48\textwidth]{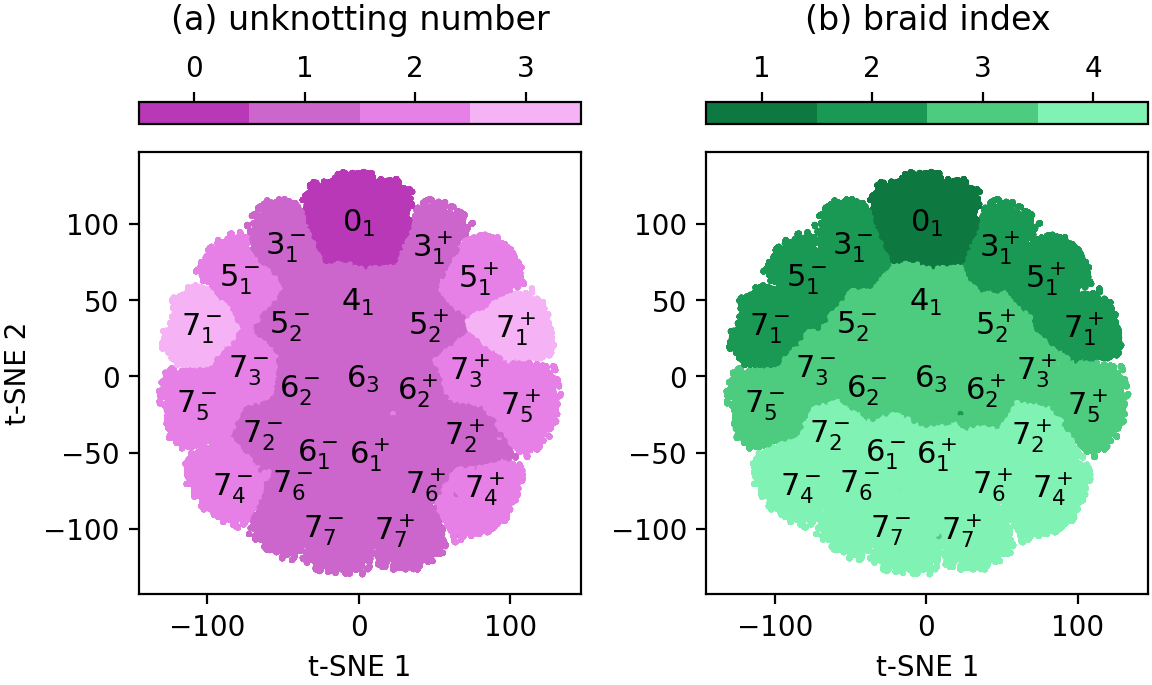}
 \caption{As in Fig.~\ref{fig:tSNE3} but highlighting (a) the unknotting number of knots and (b) their braid index.
} 
 \label{fig:tSNE4}
\end{figure}

Figure~\ref{fig:tSNE4} shows that the VAEC has also been organizing knots in the latent space by generating gradients of their unknotting number (it increases with the distance from the vertical axis, as shown in Fig.~\ref{fig:tSNE4}(a)) and the braid index, which grows from top to bottom, see Fig.~\ref{fig:tSNE4}(b). Both indices quantify knot complexity in different ways. 

The unknotting number is the minimum number of strand passages required to transform a knot into the unknot ($0_1$). For example, the knot $5_1$, a first strand passage may transform it to $3_1$, and the second to $0_1$. Hence, its unknotting number $2$, larger than the minimum non-zero value, tells us that it is not the simplest knot to morph to $0_1$. 

%%%%%%%%%%%%%%%%%%%%%%%%%%%%%%%%%%%%%%%%%
%%%%%%%%%%%%%%%%%%%%%%%%%%%%%%%%%%%%%%%%%

%%%%%%%%%%%%%%%%%%%%%%%%%%%%%%%%%%%%%%%%%
 \begin{figure}[t!]
 \centering
 \includegraphics[width=.4\textwidth]{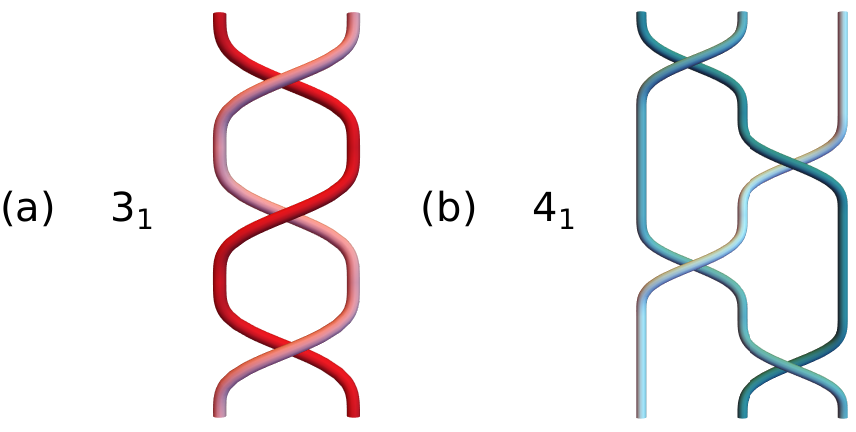}
 \caption{Braid representation for knot (a) $3_1$ with braid index $2$ and (b) $4_1$ with index $3$. The braid index corresponds to the number of vertical lanes in the braid. The knot emerges by joining the two ends (top and bottom) in each lane.
} 
 \label{fig:braid}
\end{figure} 
%%%%%%%%%%%%%%%%%%%%%%%%%%%%%%%%%%%%%%%%%

Quite simply speaking, the braid index is the minimum number of parallel lanes required to host the knot, as depicted in Fig.~\ref{fig:braid}. Torus knots such as $5_1$ only require two lanes and thus have a relatively simple braiding representation (see that of knot $3_1$ in Fig.~\ref{fig:braid}(a)). The example with knot $5_1$, with unknotting number larger than $1$ but braid index $1$, shows that these two indicators quantify different degrees of knot complexity. An opposite example is the $4_1$ knot, whose braid index $3$ (see Fig.~\ref{fig:braid}(b)) is not the smallest among nontrivial knots, and yet its unknotting number is $1$, that is, the minimum possible value.

In summary, remarkably, the VAEC, only from analyzing three-dimensional configurations labeled arbitrarily with classes, has learned a fascinating latent representation of knots: it sorts knots according to families and independent indices of complexity.

\subsection{Distinguishing among $9_{42}^\pm$ and $10_{71}^\pm$}

A more difficult test to pass concerns the recognition, within geometrically complex configurations, of knots on which training was not performed; in other words, knots unknown to the VAEC.
In particular, we check the VAEC latent representation for both chiralities of two specific chiral knots, $9_{42}$ and $10_{71}$. 
We choose these two knots because they are known to be unrecognizable by the most powerful topological invariants, such as the Jones and HOMFLY polynomials.

\begin{figure}[t!]
 \centering
 \includegraphics[width=.35\textwidth]{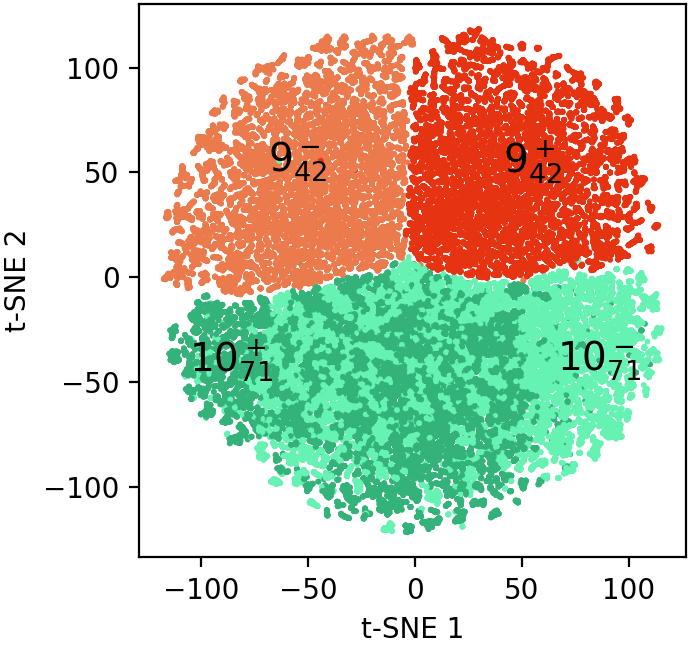}
 \caption{Embedding to two dimensions of the latent representation of the two chiral versions for knots $9_{42}$ and $10_{71}$, using the VAEC trained on chiral knots up to 7 crossings. One can easily distiguish $9_{42}^-$ from $9_{42}^+$.} 
 \label{fig:942-1071}
\end{figure}

In Fig.~\ref{fig:942-1071}, we show the latent space representation of these two knots built by the VAEC trained by the knots in Table~\ref{fig:table}, including information on their chirality: one can observe a neat separation between the knot types $9_{42}$ and $10_{71}$, irrespective of their chirality and a surprising partition by chirality of the $9_{42}$. Although the corresponding partition for $10_{71}$ is less neat, this result confirms the ability of the VAEC to capture topological features of close random curves and use them to recognize knots that are challenging even to powerful tools such as polynomial invariants.

%%%%%%%%%%%%%%%%%%%%%%%%%%%%%%%%%%%%%%%%%
%%%%%%%%%%%%%%%%%%%%%%%%%%%%%%%%%%%%%%%%%

\section{Conclusions}

In this work, we trained and tested a variational autoencoder with a classifier (VAEC) on knotted flexible ring configurations under spherical confinement. Due to confinement, the sampled configurations are highly geometrically entangled, making the identification of the underlying knot types extremely challenging for any method known in the literature. By examining the latent space representation of the VAEC, we discovered that the method can capture essential features of ring topology while filtering out noise and irrelevant details. 

In particular,  our findings indicate that knotted rings with knot types belonging to the same family (achiral, torus, or twist knots) are often closely positioned within the latent space. If knots are also classified according to their chirality, the latent space exhibits a symmetry in the position of knots with opposite chirality, with achiral knots forming an axis of symmetry. In particular, the VAEC, once trained on the set of chiral knots up to $7$ crossings,  outperforms invariants such as the Jones and HOMFLY polynomials in recognizing the chirality of mutant knots such as the $9_{42}$ and $10_{71}$ knots. Furthermore, the latent representation of the VAEC follows an interesting sorting along independent axes of knots according to their unknotting number and their braid index.

These results indicate that the designed VAEC can faithfully grasp topological concepts such as chirality, knot-type similarity, and knot complexity only by analyzing the three-dimensional coordinates of geometrically entangled configuration.
Notably, the performance of knot classification based on the latent space representation remains quite strong even for polymer chains that are longer than those used during training. This ability to generalize of the discriminative model reading the VAE latent space is likely enhanced by the necessary ability required by the generative part (the encoder-decoder main body of the VAE) to "understand" the data so that a good reconstruction of much smoother knotted rings is possible.

\acknowledgments{
Funding from research grants `ORLA$\_$BIRD2020$\_$01' and 'BAIE$\_$BIRD2021$\_$01' of Universit\`{a} degli Studi di Padova is gratefully acknowledged.  Figures~\ref{fig:table} and~\ref{fig:braid} were generated with Mathematica.}

%\bibliography{knots}


\end{document}